\documentstyle[12pt,epsfig]{article}
\newcommand{\be}{\begin{equation}}
\newcommand{\ee}{\end{equation}}
\newcommand{\ba}{\begin{eqnarray}}
\newcommand{\ea}{\end{eqnarray}}

\newcommand{\cao}{{\cal O}} 
\newcommand{\dg}{^{\dagger}}

\newcommand{\dfp}{[\mbox{d}\bar{\psi}'\mbox{d}\psi']}
\newcommand{\di}{\mbox{\scriptsize d}\,} 
\newcommand{\e}{\mbox{e}}
\newcommand{\f}{{\mbox{\tiny f}}} 
\newcommand{\ga}{\gamma_5}
\newcommand{\Id}{\mbox{1\hspace{-0.98mm}l}}   
\newcommand{\la}{\lambda}
\newcommand{\mb}[1]{\quad\mbox{ #1 }\quad}

\newcommand{\re}[1]{(\ref{#1})}

\begin{document}
\renewcommand{\baselinestretch}{1.1} \small\normalsize

\vspace*{.9cm}

\hfill {\sc HU-EP}-00/53

\vspace*{1.2cm}

\begin{center}

{\Large \bf General Ginsparg-Wilson fermions and index} 

\vspace*{0.9cm}

{\bf Werner Kerler}

\vspace*{0.3cm}

{\sl Institut f\"ur Physik, Humboldt-Universit\"at, D-10115 Berlin, 
Germany}
\hspace{3.6mm}

\end{center}

\vspace*{1.2cm}

\begin{abstract}
We show rigorously that for general Ginsparg-Wilson fermions the dimensions of 
the geometric eigenspace and of the algebraic one for zero modes agree so that 
the index theorem on the lattice is not spoiled by unwanted additional terms.
\end{abstract}

\vspace*{1.2cm}

In recent years the formulation of the index theorem on lattices has become
possible \cite{na94,ha98,lu98}. It has been shown \cite{ne98} that the Dirac 
operator $D$ in \cite{na94} obeys the form $\{\ga,D\}= 2 D \ga D$ of the
Ginsparg-Wilson (GW) relation \cite{gi82}. This form has also been assumed
in \cite{lu98}. On the other hand, in \cite{ha98} the index theorem has been
claimed to hold for $D$ satisfying the more general GW relation \cite{gi82}
\be
\{\ga,D\}= 2 D \ga R D 
\label{aGW}
\ee
where $R$ is a hermitean operator which commutes with $\gamma$-matrices. The 
derivation given there relies on the eigenvectors, i.e.~is solely based on the 
consideration of the geometric eigenspace. However, in general the dimension 
of the geometric eigenspace can be smaller than that of the algebraic 
eigenspace, in which case unwanted additional terms would spoil the index 
theorem. In view of the many works relying on the general form \re{aGW}, it 
appears important to clarify whether in that case such a difference occurs. In 
the present letter we find that settling this question requires a detailed
study of properties of the eigennilpotents and show rigorously that the 
respective dimensions for zero modes agree.

The index theorem follows from the global chiral Ward identity. This identity
is obtained requiring $\frac{\di}{\di\eta} \int\dfp \e^{-S_\f'} \cao' 
\big|_{\eta=0} = 0$ for the transformation $\psi'= \exp(i\eta\ga)\psi$, 
$\bar{\psi}'= \bar{\psi} \exp(i\eta\ga)$. Since we have to deal with the case 
where the Dirac operator $D$ has zero modes, we must make sure to account 
properly for them. We therefore replace $D$ by $D-\zeta$ with the parameter 
$\zeta$ being in the resolvent set (i.e.~not in the spectrum of $D$) allowing 
$\zeta$ to go to zero only in the final result. With  $S_\f'=
\bar{\psi'}(D-\zeta)\psi'$ we thus obtain
\be
\frac{1}{2}\mbox{Tr}\Big((D-\zeta)^{-1}\{\ga,D\}\Big)-
\zeta\,\mbox{Tr}\Big((D-\zeta)^{-1}\ga\Big)=0 \;.
\label{gwa2}
\ee
Of course, the validity of the identity \re{gwa2} can also be verified 
directly. Obviously it is just a particular decomposition of 
$\mbox{Tr}\,\ga=0$.

In order to evaluate \re{gwa2} we use the fact that the resolvent of $D$ is 
given by \cite{ka66}
\be
(D-\zeta)^{-1}=-\sum_{j=1}^s\Big((\zeta-\la_j)^{-1}P_j+\sum_{k=1}^{d_j-1}
(\zeta-\la_j)^{-k-1}Q_j^{\,k}\Big)\;.
\label{reso}
\ee 
The operators $P_j$ and $Q_j$ in \re{reso} satisfy $P_j P_l=\delta_{jl}P_j$, 
$P_j Q_l=Q_l P_j=\delta_{jl}Q_j$, and $Q_j Q_l=0$ for $j\ne l$. The $P_j$ 
project on the algebraic eigenspaces $M_j$ with dimensions $d_j=
\mbox{ Tr }P_j$. The $Q_j$ have the property $Q_j^{\,d_j}=0$, i.e.~they are 
nilpotents. In terms of these operators the spectral representation of $D$ 
becomes \cite{ka66}
\be
D=\sum_j(\la_j P_j+Q_j)\;.
\label{spec}
\ee 
The direct sum of the spaces $M_j$ makes up the total space. We emphasize that 
they in general are not orthogonal. Correspondingly then in general $P_j$, 
which projects on $M_j$ along $\tilde{M_j}=M_1\bigoplus\ldots\bigoplus 
M_{j-1}\bigoplus M_{j+1}\bigoplus\ldots\bigoplus M_s$, is also not orthogonal
and one has $P_j\dg\ne P_j$ since $\tilde{M_j}$ is not the orthogonal 
complement of $M_j$. Therefore in the following we have to work carefully with 
general projections and subspaces.

By inserting \re{aGW} into the identity \re{gwa2} we obtain
\be
\mbox{Tr}(\ga RD)-\zeta\,\mbox{Tr}\Big(\ga(D-\zeta)^{-1}\Big)
                 +\zeta^2\,\mbox{Tr}\Big(\ga R(D-\zeta)^{-1}\Big)=0\;.
\label{e2}
\ee
Dividing \re{e2} by $\zeta$, expressing $(D-\zeta)^{-1}$ by \re{reso}, and 
integrating over $\zeta$ around a circle enclosing only the eigenvalue 
$\la_k=0$ we find 
\be
\mbox{Tr}(\ga RD)+\mbox{Tr}\Big(\ga(P_k+RQ_k)\Big)=0 \mb{for} \la_k=0\;.
\label{e2i}
\ee 

To evaluate \re{e2i} further we need more information about the nilpotents 
$Q_j$. They account for the fact that for $d_j>1$ the dimension $g_j$ of the 
geometric eigenspace can be smaller than the dimension $d_j$ of the 
algebraic one. The dimension $d_j$ equals the multiplicity of the 
solution $\la=\la_j$ of $\mbox{det}(D-\la\Id)=0$ while $g_j$ is given by 
$g_j=\mbox{ dim ker }(D-\la_j\Id)$. From the latter relation and \re{spec} we 
obtain 
\be
\mbox{ rank }Q_j= d_j-g_j \;.
\label{rank}
\ee
Furthermore, since $\mbox{det}(D-\la_j\Id)=0$ one has $g_j=  
\mbox{ dim ker }(D-\la_j\Id) \ge1$ and therefore $\mbox{rank }Q_j\le d_j-1$ 
and $1\le g_j\le d_j$.      

To specify the spaces in more detail we note that the linearly independent set 
of eigenvectors $f_{jl}$ with $Df_{jl}=\la_jf_{jl}$ and $l=1,\ldots,g_j$ spans
the geometric eigenspace $M_j'$ of $D$ with dimension $g_j$ while $P_j$ as
introduced above projects on the algebraic one $M_j$ with dimension $d_j$. For 
$g_j< d_j$ we choose $M_j''$ such that $M_j=M_j'\bigoplus M_j''$. This 
defines projections $P_j'$ and $P_j''$ where $P_j'$ projects on $M_j'$ along 
$M_j''\bigoplus \tilde{M}_j$ and $P_j''$ on $M_j''$ along $M_j'\bigoplus 
\tilde{M_j}$. One then has 
\be
P_j=P_j'+P_j''
\label{p'}
\ee
with $P_j'P_j''=P_j''P_j'=0$ and $\mbox{Tr } P_j'=g_j$. 

It will be important that with this decomposition according to \re{spec} we 
have
\be
Q_j\phi=0 \quad{\rm for} \quad \phi\in M_j'
\label{Qf}
\ee 
characteristic for the geometric eigenspace. In terms of operators this means
that
\be
Q_jP_j'=0\quad\mbox{ and }\quad Q_kP_j''=Q_j\;.
\label{Qp}
\ee 
On the other hand, for the products $P_j'Q_j$ and $P_j''Q_j$ in general various
results can occur which only have to satisfy 
\be
P_j'Q_j+P_j''Q_j=Q_j
\label{Qpp}
\ee      
as is needed because of $P_jQ_j=Q_j$.

We next impose the usual requirement of $\ga$-hermiticity of $D$,
\be
D\dg = \ga D \ga\;.
\label{ga}
\ee
Then the GW relation \re{aGW} can be written as $D+D\dg=2D\dg RD$ which implies
that 
\be
Df_{kl}=0\quad\mbox{ and }\quad D\dg f_{kl}=0\quad\mbox{ for }\quad\la_k=0
\label{DD}
\ee 
hold simultaneously.

 From \re{DD} and \re{ga} one obtains $[\ga,D]f_{kl}=0$ for $\la_k=0$. 
Therefore the $f_{kl}$ can be chosen such that $\ga f_{kl}=s_{kl}f_{kl}$ with 
$s_{kl}= \pm1$. The 
sets of $f_{kl}$ with $s_{kl}=+1$ and $s_{kl}=-1$ then define subspaces 
$M_k'^{(+)}$ and $M_k'^{(-)}$, respectively, with $M_k'=M_k'^{(+)}\bigoplus 
M_k'^{(-)}$. This in turn defines the projections $P_k'^{(+)}$ on 
$M_k'^{(+)}$ along $M_k'^{(-)}\bigoplus M_k''\bigoplus \tilde{M}_k$ and 
$P_k'^{(-)}$ on $M_k'^{(-)}$ along $M_k'^{(+)}\bigoplus M_k''\bigoplus 
\tilde{M}$. We thus get
\be
P_k'=P_k'^{(+)}+P_k'^{(-)}
\label{p+}
\ee
with $P_k'^{(+)}P_k'^{(-)}=P_k'^{(-)}P_k'^{(+)}=0$ and 
$\mbox{Tr }P_k'^{(\pm)}=g_k^{(\pm)}$, where $g_k^{(\pm)}$ denotes the numbers 
of modes with $s_{kl}=\pm1$, respectively, and $g_k=g_k^{(+)}+g_k^{(-)}$. 

By inserting \re{p'} and \re{p+} into \re{e2i} we now obtain the more detailed
form
\be
\mbox{Tr}(\ga RD)+g_k^{(+)}-g_k^{(-)}+\mbox{Tr}(\ga P_k'')+
\mbox{Tr}(\ga RQ_k)=0 \mb{for} \la_k=0\;.
\label{e2k}
\ee 
Comparing \re{e2k} with the relation given in \cite{ha98} we see that the 
terms 
\be
+\mbox{Tr}(\ga P_k'')+\mbox{Tr}(\ga RQ_k)
\label{e2t}
\ee
are missing there. The occurrence of the terms \re{e2t} is related to the 
possibility that the dimension of the geometric eigenspace can be smaller than 
that of the algebraic one. In fact, according to \re{rank} we have 
$\mbox{ rank }Q_k=d_k-g_k$ and from \re{p'} it follows that $\mbox{Tr }P_k''= 
d_k-g_k$. Thus it becomes obvious that it is $g_k=d_k$ which is needed
to make \re{e2t} vanish. This means that, in order to get rid of the unwanted 
terms \re{e2t}, $D$ should have a property which guarantees that the solutions 
of the eigenvalue problem satisfy $g_k=d_k$.

It can be shown \cite{ka66} that if $D$ is normal, $[D,D\dg]=0$, one has 
$Q_j=0$ for all $j$. Thus, for such $D$ one gets $g_j=d_j$ for all $j$ and, in 
particular, $g_k=d_k$ for $\la_k=0$. In addition with normality of $D$ 
one has \cite{ka66} $P_j\dg=P_j$ for all $j$, 
i.e.~orthogonality of the eigenprojections and of the associated subspaces.
However, a Dirac operator satisfying \re{aGW} is in general not normal.
To make this explicit we note that from \re{aGW} using \re{ga} and $[\ga,R]=0$ 
one obtains $[D,D\dg]=2D\dg[R,D]D\dg$. Thus it is seen that one would need 
$[R,D]=0$ to make $D$ normal, which to fulfil generally would require $R$ to 
be a multiple of the identity, i.e.~to restrict to the simple form 
$\{\ga,D\}= 2 D \ga D$ of the GW relation.

To proceed with the more general relation \re{aGW} we note that $Q_k=0$ is 
actually needed for $\la_k=0$ only. In the following we prove this weaker 
property.   For this purpose we first observe that from \re{DD} with \re{Qf} 
we have 
\be
Q_kP_k'=0 \quad{\rm and}\quad Q_k\dg P_k'=0   \mb{for} \la_k=0\;.
\label{Qff}
\ee 
We next remember that $P_k'$ projects on $M_k'$ along $M_k''\bigoplus
\tilde{M}_k$ so that by definition $(P_k')\dg$ projects on $(M_k''\bigoplus 
\tilde{M}_k)^{\perp}$ along $M_k'^{\perp}$ where $\perp$ denotes orthogonal 
complements. On the other hand, from \re{DD} it follows that $P_k'$ and 
$(P_k')\dg$ both project on $M_k'$. Therefore we get 
$M_k'=(M_k''\bigoplus \tilde{M}_k)^{\perp} $ which implies 
\be
(P_k')\dg=P_k'    \mb{for} \la_k=0\;.
\label{PdP}
\ee 
 From this and the adjoint of the second relation in \re{Qff} we now obtain 
$P_k'Q_k=0$ and thus, with \re{Qpp}, arrive at the relations
\be
P_k'Q_k=0 \quad{\rm and}\quad P_k''Q_k=Q_k   \mb{for} \la_k=0\;.
\label{PQ}
\ee
Obviously \re{PQ} can be satisfied with $Q_k=0$. This is, however, not possible
with $Q_k\ne0$, as we shall show below, which will complete our proof.

In case of $Q_j\ne0$, for some integer $n$ with $1\le n\le d_j-1$ we have 
$Q_j^{n+1}=0$ but $Q_j^n\ne0$. Then, since the range of $Q_j^n$ is nonzero, 
there is a nonzero vector $\phi$ which satisfies $Q_j\phi=0$. Further, there 
is some vector $\psi_1$ such that $\phi=Q_j^n\psi_1$. For $n>1$ we can 
successively define
\be
\psi_{\nu}=Q_j\psi_{\nu-1} \quad{\rm for}\quad \nu=2,\ldots,n
\label{ps}
\ee
and we get $\phi=Q_j\psi_n$ for $n\ge1$. Decomposing $\psi_{\nu}$
according to
\be
\psi_{\nu}=\varphi_{\nu}+\chi_{\nu}
\quad\mbox{with}\quad \varphi_{\nu}\in M_j'\;,\; \chi_{\nu}\in M_j''
\ee
and noting that by \re{Qf} $Q_j\varphi_{\nu}=0$ these relations become
\be
\chi_{\nu}=Q_j\chi_{\nu-1} \quad{\rm with}\quad \nu=2,\ldots,n \quad{\rm for}
\quad n>1
\label{ch}
\ee
and $\phi=Q_j\chi_n$ for $n\ge1$. Because of $Q_j\phi=0$ by \re{Qf} we have 
$\phi\in M_j'$. Thus it is seen that the nilpotent property of $Q_j$
relies on a sequence of $n-1$ transformations within $M_j''$, one from $M_j''$
to $M_j'$, and a final one which then according to \re{Qf} gives zero. 
Obviously it is crucial for this property that the indicated mapping from 
$M_j''$ to $M_j'$ is possible.
 
The $\chi_{\mu}$ with $\mu=1,\ldots,n$ for $n=r_j$ provide a basis of $M_j''$ 
and for $n<r_j$ a basis of a $n$-dimensional subspace of $M_j''$. In the latter 
case we repeat the above procedure for the remaining $(r_j-n)$--dimensional 
subspace of $M_j''$ in which $Q_j^{n_2+1}=0$ but $Q_j^{n_2}\ne0$ for some 
integer $n_2$ with $1\le n_2\le n_1\equiv n$. Possibly further repetitions are 
needed until the space $M_j''$ is exhausted and we reach $n_1+n_2+\ldots+n_h
=r$ with $n_1\ge n_2\ge \ldots\ge n_h\ge1$ and $1\le h\le g_j$. Clearly in each
of the subspaces of $M_j''$ involved in this process it remains crucial that 
$M_j'$ can be reached by the transformations in the indicated way.

Because the range of $P_k''$ is $M_k''$, the condition $P_k''Q_k=Q_k$ for 
$\la_k=0$ of \re{PQ} implies that the range of $Q_k$ must be within $M_k''$. 
Therefore it cannot map to $M_k'$ as we have shown above to be necessary for 
the nilpotent property of $Q_k\ne0$. Thus $Q_k\ne0$ is excluded and \re{PQ} 
is indeed only satisfied by $Q_k=0$, which completes the proof.

With $Q_k=0$ by \re{rank} we now have $g_k=d_k$ and $P_k''=0$ for $\la_k=0$ and 
arrive at the result that the terms \re{e2t} vanish. Further, because of $P_k'=
P_k$, as for $M_k'$ before, the decomposition $M_k=M_k^{(+)} \bigoplus 
M_k^{(-)}$ with the projections $P_k^{(+)}$ and $P_k^{(-)}$ can be introduced
and one sees that $g_k^{(\pm)}= d_k^{(\pm)}$ holds. With $P_k'=P_k$ from 
\re{PdP} in addition $P_k\dg=P_k$ for $\la_k=0$ gets obvious.

\section*{Acknowledgement}

\hspace{3mm}
I wish to thank Michael M\"uller-Preussker and his group for their kind 
hospitality.

\end{document}